\documentclass[runningheads,orivec]{llncs}
\usepackage{amsmath}
\usepackage{booktabs}
\usepackage[T1]{fontenc}
\usepackage{graphicx}
\usepackage{marvosym}
\usepackage{ifsym}
\usepackage{hyperref}
\usepackage{color}
\hypersetup{
    colorlinks=true,
    linkcolor=blue,
    citecolor=blue,
    filecolor=blue,
    urlcolor=blue
}

\usepackage{array}
\usepackage{multirow}

\begin{document}
\title{Multi-modal brain MRI synthesis based on SwinUNETR}
%
%\titlerunning{Abbreviated paper title}
% If the paper title is too long for the running head, you can set
% an abbreviated paper title here
%
\author{Haowen Pang \and Weiyan Guo \and Chuyang Ye$^{(\textrm{\Letter})}$}

\authorrunning{H. Pang et al.}
% First names are abbreviated in the running head.
% If there are more than two authors, 'et al.' is used.
%
\institute{School of Integrated Circuits and Electronics, Beijing Institute of Technology, Beijing, China.\\
\email{chuyang.ye@bit.edu.cn}}
\maketitle              % typeset the header of the contribution
\begin{abstract}
Multi-modal brain magnetic resonance imaging (MRI) plays a crucial role in clinical diagnostics by providing complementary information across different imaging modalities.
However, a common challenge in clinical practice is missing MRI modalities.
In this paper, we apply SwinUNETR to the synthesize of missing modalities in brain MRI.
SwinUNETR is a novel neural network architecture designed for medical image analysis, integrating the strengths of Swin Transformer and convolutional neural networks (CNNs).
The Swin Transformer, a variant of the Vision Transformer (ViT), incorporates hierarchical feature extraction and window-based self-attention mechanisms, enabling it to capture both local and global contextual information effectively.
By combining the Swin Transformer with CNNs, SwinUNETR merges global context awareness with detailed spatial resolution.
This hybrid approach addresses the challenges posed by the varying modality characteristics and complex brain structures, facilitating the generation of accurate and realistic synthetic images.
We evaluate the performance of SwinUNETR on brain MRI datasets  and demonstrate its superior capability in generating clinically valuable images.
Our results show significant improvements in image quality, anatomical consistency, and diagnostic value.
% 150--250 words.
\end{abstract}

\section{Introduction}
Brain tumors are a serious brain disease.
The incidence of brain tumors varies globally and is a significant cause of mortality, especially in cases of delayed diagnosis and treatment.
The symptoms of brain tumors are diverse and require treatments tailored to the size and location of the tumor.
Therefore, early and accurate diagnosis is crucial for developing an effective treatment plan.
Magnetic resonance imaging (MRI) has become an essential tool for brain tumor diagnosis due to its superior soft tissue contrast and ability to provide detailed anatomical information.
Multi-modal MRI can enhance diagnostic accuracy by providing additional information about tumor characteristics.
Thus, multi-modal MRI plays a critical role in the comprehensive assessment of brain tumors, from diagnosis to treatment planning and monitoring.

\begin{figure}[!t]
\includegraphics[width=\textwidth]{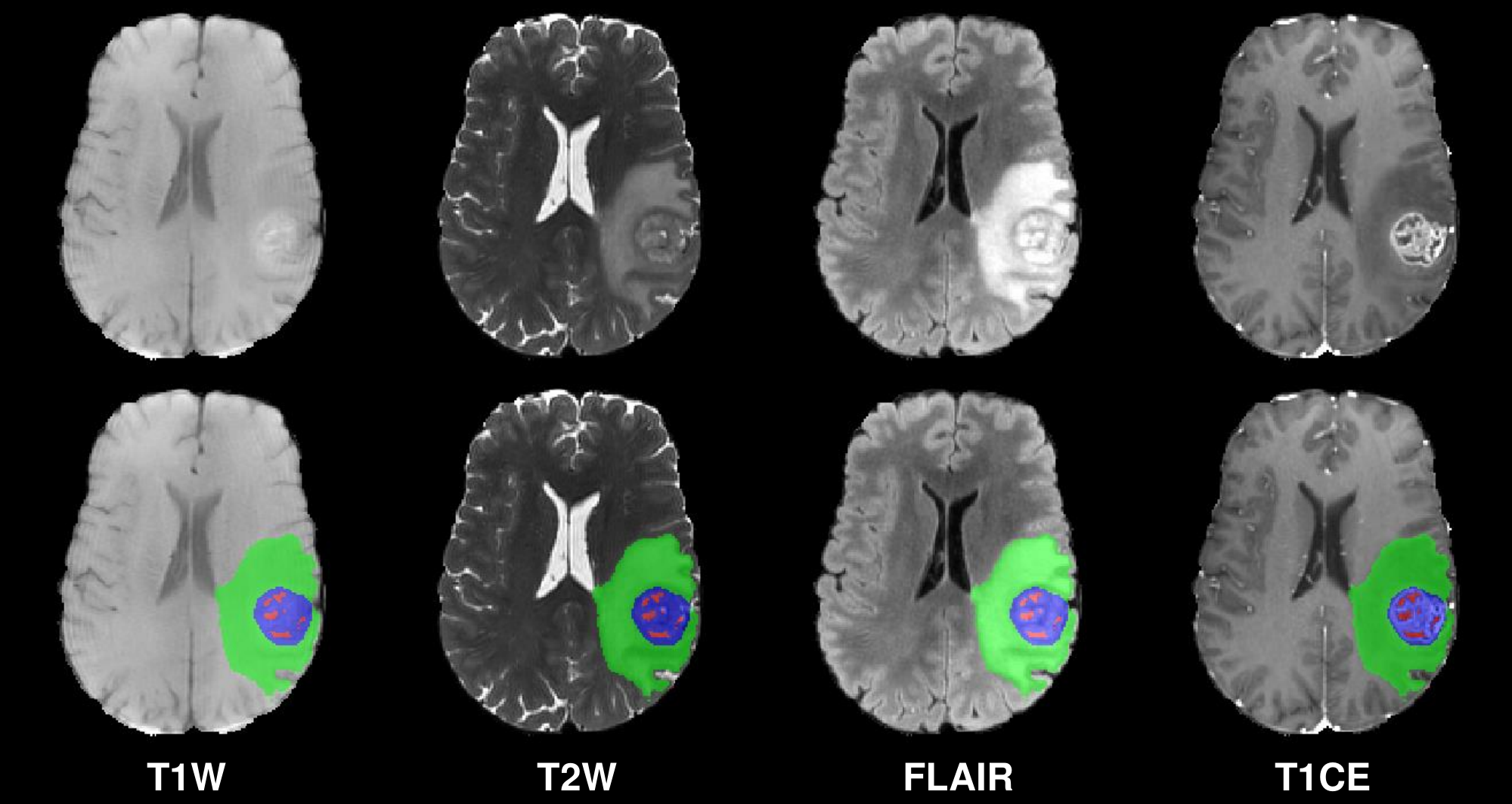}
\caption{Multi-modal MRI plays a critical role in comprehensive assessment of brain tumors.} \label{fig1}
\end{figure}

As shown in Fig.~\ref{fig1}, T1-weighted images (T1w) provide high-resolution anatomical details, crucial for identifying brain tissue and tumor structures, as well as detecting hemorrhage and fat within the tumor.
T2-weighted images (T2w) are highly sensitive to water content, making them ideal for detecting edema and depicting the overall tumor morphology.
Fluid-attenuated inversion recovery (FLAIR) imaging suppresses cerebrospinal fluid signals, improving the visibility of lesions near the ventricles and cortical surfaces.
Post-contrast T1-weighted imaging (T1CE) highlights tumors with blood-brain barrier disruption, aiding in differentiating tumor types and detecting metastasis and recurrence.
The combination of multi-modal MRI provides a comprehensive assessment of brain tumors, enhancing diagnostic accuracy and informing treatment strategies.
However, a common challenge in clinical practice is missing MRI modalities, due to time constraints or image artifacts
Therefore, synthesizing the missing MRI modality from available modalities is of significant importance.

In recent years, deep learning-based model have been widely used in MRI synthesis.
Dar et al. \cite{dar2019image} propose pGAN and cGAN for multi-contrast MRI synthesis.
Dalmaz et al. \cite{dalmaz2022resvit} propose ResViT, which combines the contextual sensitivity of vision transformers with the precision of convolution operators and the realism of adversarial learning.
Liu et al. \cite{liu2023one} formulate missing MRI synthesis as a sequence-to-sequence learning problem and propose a multi-contrast multi-scale Transformer (MMT) that can take any subset of input contrasts and synthesize the missing ones.
Meng et al. \cite{meng2024multi} propose a unified Multi-modal Modality-masked Diffusion Network (M2DN) based on the diffusion model\cite{ho2020denoising}.
However, most of these methods are highly complex.
In this work, we use a very simple method based on a encoder-decoder network to synthesize MRI, which provides a baseline for future research.

\section{Methods}

\subsection{Network structure}
\begin{figure}[!t]
\includegraphics[width=\textwidth]{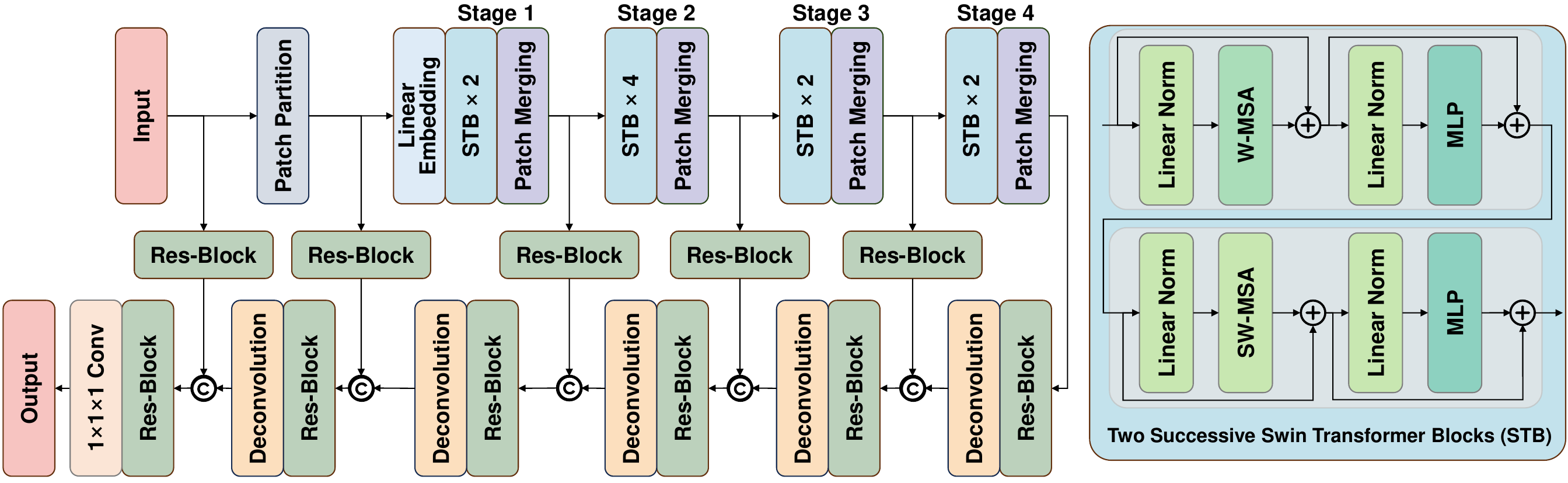}
\caption{The architecture of SwinUNETR.} \label{fig2}
\end{figure}

In this work, we apply SwinUNETR~\cite{hatamizadeh2021swin,tang2022self} to synthesize missing modalities in brain MRI. 
SwinUNETR utilizes a U-shaped architecture, with the Swin Transformer~\cite{liu2021swin} serving as the encoder and a CNN-based decoder connected via skip connections at multiple resolutions.
Our implementation of SwinUNETR is shown in Fig.~\ref{fig2}.

We did not modify the network structure of SwinUNETR.
For a detailed description of the SwinUNETR architecture, please refer to \cite{hatamizadeh2021swin} and \cite{tang2022self}.
The only change we made is setting the number of input channels in the network to 3 and the number of output channels to 1.

\subsection{Data preprocessing}

During the training phase, we standardized each modality using the Z-score:
\begin{equation}
\label{eq:Z-Score}
z_{i}=\frac{x_i-\mu}{\sigma}
\end{equation}
where $\mu$ and $\sigma$ represent the mean and standard of image voxel values, respectively, $x_{i}$ is the original voxel value, and $z_{i}$ is the standardized voxel value.

\subsection{Implementation Details}

For the four modalities, T1w, T2w, FLAIR, and T1CE, we trained a SwinUNETR-based image synthesis model for each missing modality scenario.
In total, four SwinUNETR models were trained, each addressing the absence of a specific modality.

For example, when T1CE is missing, we trained a SwinUNETR model with a three-channel input and a single-channel output.
The input modalities include T1w, T2w, and FLAIR, and the model is supervised using the real T1CE modality with \textit{Mean Squared Error} (MSE) loss.
\begin{equation}
\label{eq:MSE}
\text{MSE} = \frac{1}{n} \sum_{i=1}^{n} (y_i - \hat{y}_i)^2
\end{equation}
where $y_i$ denotes the voxel value of the real image, $\hat{y}_i$ represents the predicted voxel value, and $n$ denotes the number of voxels.

To save memory, we employed a patch-based training strategy.
Specifically, during training, we randomly extracted patches of size $128\times128\times128$ from the images.
In testing, we used a sliding window approach for prediction, with overlapping regions fused using Gaussian weighting.

We used the 1251 image pairs provided by the challenge as the training set and 219 image pairs as the validation set~\cite{karargyris2023federated,li2023brain,baid2021rsna,menze2014multimodal,bakas2017advancing}, without using additional data or pre-trained models.
We did not use the tumor masks provided in the training set.

The batch size was set to 1, and the number of epochs was set to 100.
The Adam optimizer was used with the default learning rate.
All experiments were implemented using Python 3.9.5 and PyTorch 1.9.0.
The experiments were run on a workstation equipped with an NVIDIA Tesla V100 GPU with 32GB of memory.

\section{Results}

\begin{figure}[!t]
\includegraphics[width=\textwidth]{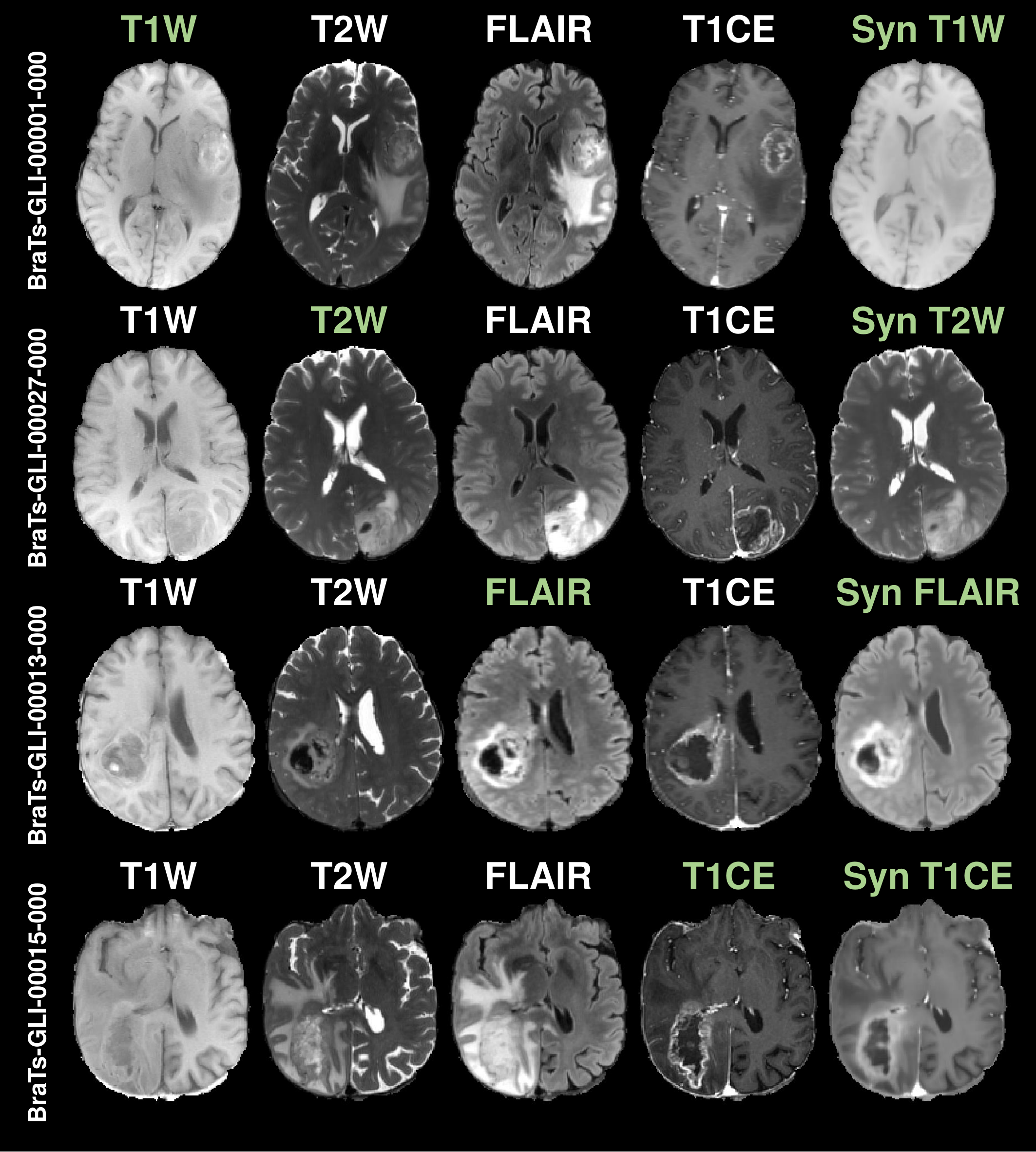}
\caption{Four examples corresponding to the missing T1w, T2w, FLAIR, and T1CE. The first four columns display the real images, while the final column shows the synthesized images generated by our method.} \label{fig3}
\end{figure}

After completing the training, we evaluated our model using the validation set provided by the challenge as the test set.
Following the "dropout\_modality.py" script supplied by the challenge, we randomly dropped one modality for each patient and performed testing.
Fig.~\ref{fig3} presents four examples corresponding to the missing T1w, T2w, FLAIR, and T1CE modalities.
The first four columns display the real images, while the final column shows the synthesized modality generated by our method.

\begin{figure}[!t]
\includegraphics[width=\textwidth]{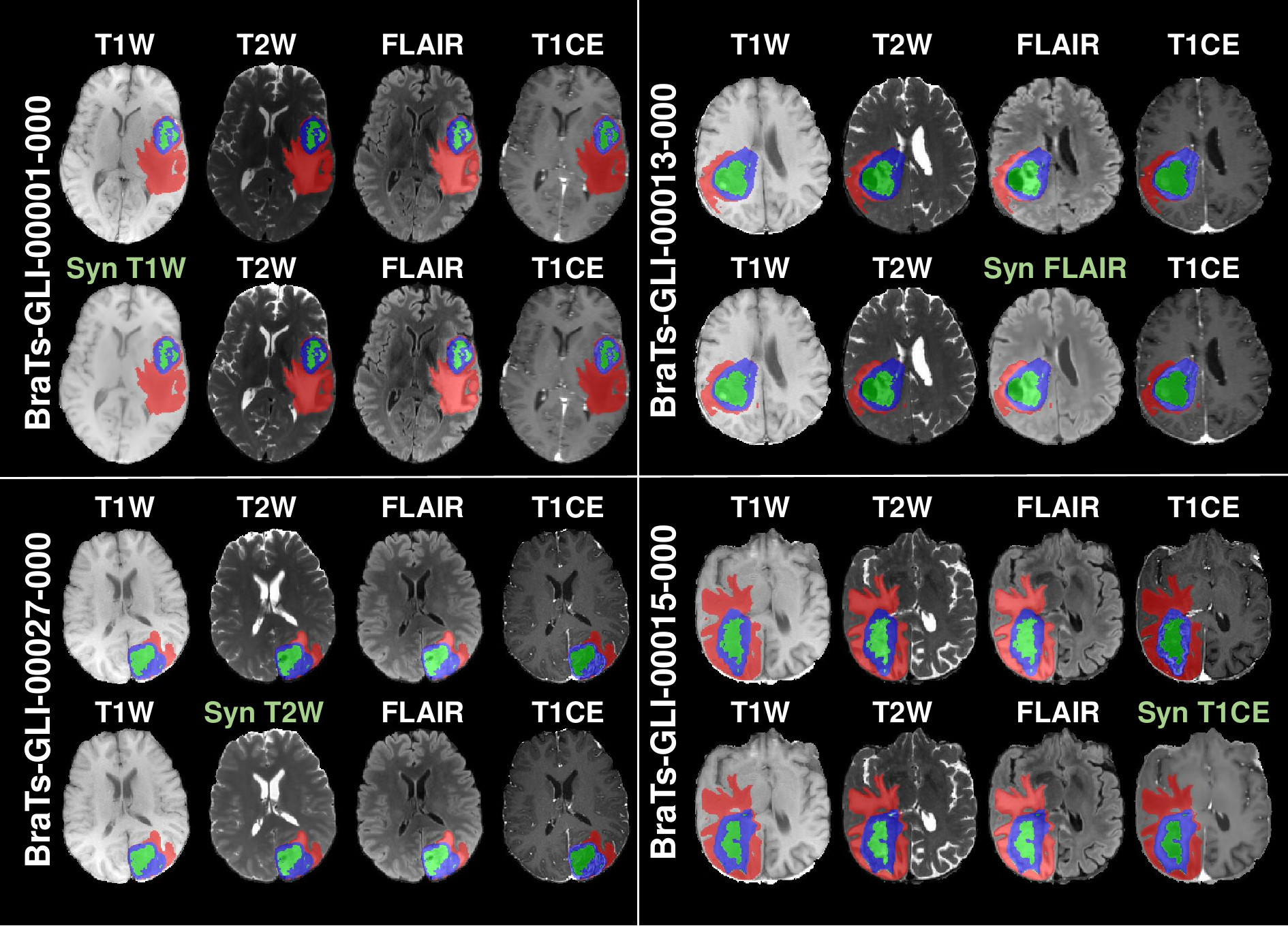}
\caption{The segmentation results of the four examples shown in Fig.~\ref{fig3}. In each case, the top row shows the segmentation results obtained using the real images, while the bottom row shows the results obtained by using one synthesized image along with the other three real images.} \label{fig4}
\end{figure}

Subsequently, we used one synthesized images along with three real images for brain tumor segmentation.
For comparison, we also tested using all four real images.
The segmentation results for the four examples shown in Fig.~\ref{fig4} are presented.
In each case, the top row displays the segmentation results obtained using the real images, while the bottom row shows the results obtained using one synthesized image combined with the other three real images.
As depicted in Fig.~\ref{fig4}, the segmentation results using the synthesized images closely resemble those obtained with the real images.

Due to the absence of tumor masks in the validation set, we calculated the Dice coefficient between the segmentation masks obtained using the synthesized images and those obtained using the real images. The results are presented in Table \ref{tab1}.

\begin{table}[!t]
\caption{Dice coefficient between the segmentation masks obtained using the synthesized images and those obtained using the real images.}\label{tab1}
\centering
\begin{tabular}{>{\centering\arraybackslash}p{3cm}>{\centering\arraybackslash}p{3cm}}
\toprule
Region&Dice $\uparrow$\\
\midrule
Label 1&0.9045$\pm$0.1006\\
Label 2&0.7795$\pm$0.3038\\
Label 3&0.8055$\pm$0.2980\\
\bottomrule
\end{tabular}
\end{table}

\begin{figure}[!t]
\includegraphics[width=\textwidth]{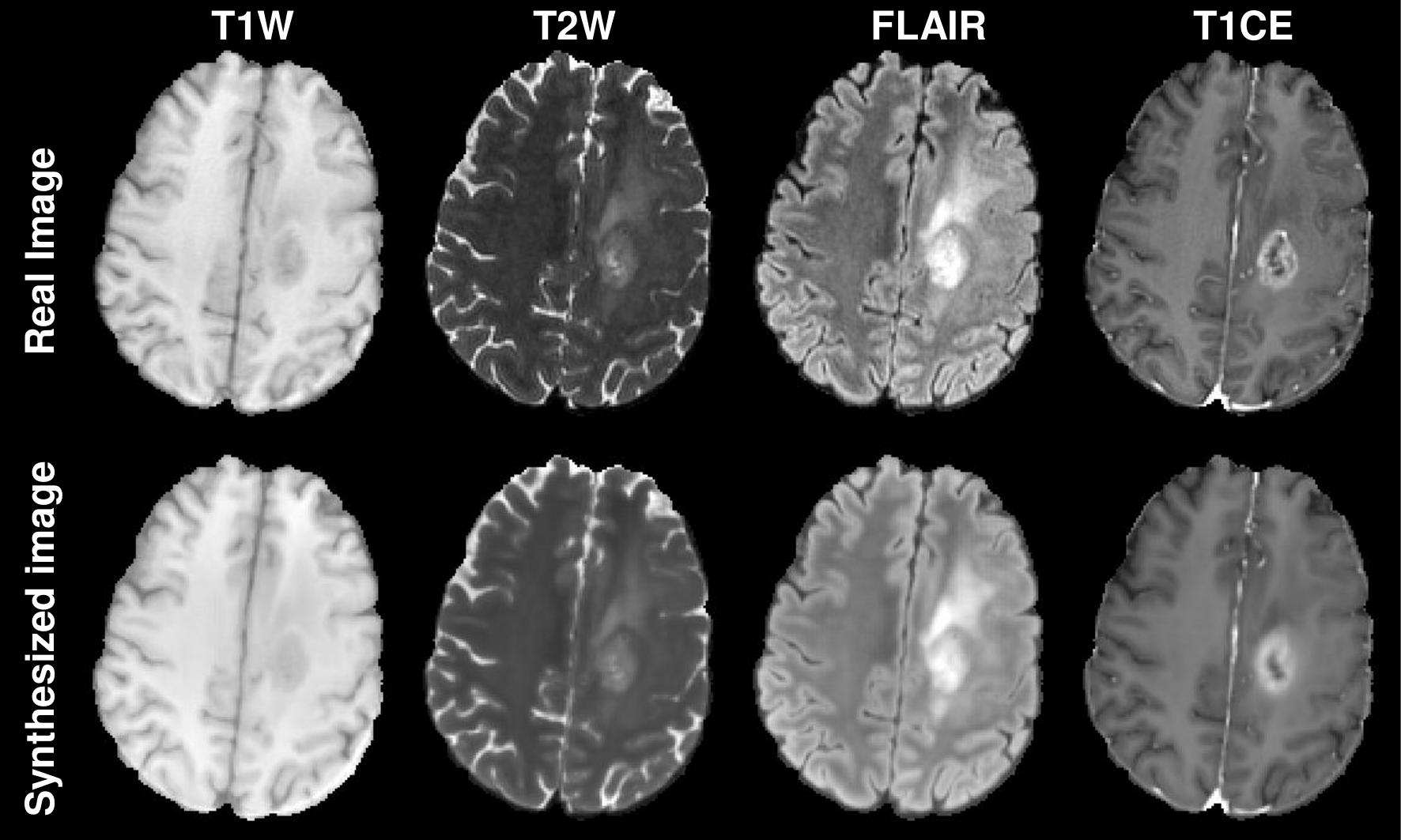}
\caption{An example of missing modality synthesis, where three modalities are used to synthesize the missing one. The first row displays the four real modalities, while the second row presents the four synthesized modalities.} \label{fig5}
\end{figure}

Additionally, we tested each patient's missing modality scenario.
Fig.~\ref{fig5} shows an example, where the first row displays the real images, and the second row presents the synthesized images.

Since the validation set did not include tumor masks, we were unable to calculate the Structural Similarity Index Measure (SSIM) for the regions inside and outside the tumors.
Instead, we calculated the SSIM for the entire image.
The results are presented in Table \ref{tab2}.
We observed that the synthesis quality of T1w images is the best, followed by T2w, while the synthesis quality of FLAIR and T1CE images was relatively lower.

\begin{table}[!t]
\caption{The SSIM between the synthesized images and the real images.}\label{tab2}
\centering
\begin{tabular}{>{\centering\arraybackslash}p{3cm}>{\centering\arraybackslash}p{3cm}}
\toprule
Missing Modality&SSIM (\%) $\uparrow$\\
\midrule
T1w&95.3784$\pm$1.6284\\
T2w&94.9089$\pm$2.4359\\
FLAIR&93.0210$\pm$2.0473\\
T1CE&93.6387$\pm$2.3147\\
\bottomrule
\end{tabular}
\end{table}

The challenge organizers evaluated our method on the test set.
The evaluation focused on general image quality and the performance of a downstream tumor segmentation algorithm applied to the synthesized image set.
The SSIM was used to quantify the realism of the synthesized images compared to clinically acquired real images.
The infilled image volume was segmented using a state-of-the-art BraTS segmentation algorithm as a downstream task, and Dice scores were calculated for three tumor structures. The FeTS algorithm~\cite{pati2022federated1,pati2022federated2} was used to evaluate the performance of the algorithms.
The results on the test dataset are presented in Table \ref{tab3}.

\begin{table}[!t]
\caption{Results on the test dataset.}\label{tab3}
\centering
\begin{tabular}{>{\centering\arraybackslash}p{1.2cm}>{\centering\arraybackslash}p{1.2cm}>{\centering\arraybackslash}p{1.4cm}>{\centering\arraybackslash}p{1.4cm}>{\centering\arraybackslash}p{1.7cm}>{\centering\arraybackslash}p{1.7cm}>{\centering\arraybackslash}p{1.7cm}}
\toprule
\multicolumn{2}{c}{Metrics} & Mean & Std & 25 Quantile & Median & 75 Quantile \\
\midrule
\multicolumn{2}{c}{SSIM} & 0.8182 & 0.0192 & 0.8050 & 0.8176 & 0.8306 \\
\midrule
\multirow{3}*{Dice}& ET & 0.6970 & 0.3088 & 0.5978 & 0.8250 & 0.9284 \\
 & TC & 0.7425 & 0.3092 & 0.6716 & 0.8873 & 0.9503 \\
 & WT & 0.8343 & 0.1984 & 0.8344 & 0.9015 & 0.9386 \\
\midrule
\multirow{3}*{HD95} & ET & 31.03 & 91.15 & 1.00 & 3.00 & 10.28 \\
 & TC & 28.47 & 81.15 & 3.16 & 6.08 & 12.08 \\
 & WT & 16.99 & 44.83  & 4.90  & 7.62  & 11.58  \\
\bottomrule
\end{tabular}
\end{table}

\section{Discussion}

In this paper, we applied SwinUNETR to the task of missing modality synthesis for brain MRI, with modifications made only to the input and output channels (set to 3 input channels and 1 output channel) to suit the task.
We trained four models to address the different missing modality scenarios.
Results on the validation set and test dataset demonstrated the effectiveness of SwinUNETR in synthesizing missing modalities.
Additionally, we used the synthesized image along with the three real images for brain tumor segmentation.
The segmentation results revealed no significant differences, suggesting that the synthesized missing modality can be effectively used in brain tumor segmentation tasks.

However, there are several limitations in our work.
First, we did not use brain tumor masks, which could provide valuable information for improving missing modality synthesis.
In future work, we plan to incorporate masks during training.
Second, our network structure lacks a discriminator.
Previous studies have shown that integrating a discriminator within an adversarial learning framework can enhance image synthesis quality.
Furthermore, diffusion models have recently achieved remarkable success in image synthesis.
In future work, we aim to explore the application of diffusion models for MRI missing modality synthesis.

%
% ---- Bibliography ----
%
% BibTeX users should specify bibliography style 'splncs04'.
% References will then be sorted and formatted in the correct style.
%
\bibliographystyle{splncs04}
% \bibliography{mybibliography}

\end{document}